\documentclass[aps,prl,reprint,showpacs,superscriptaddress,longbibliography]{revtex4-1}
\usepackage[english]{babel}
\usepackage{amsmath,amssymb,bbm,graphicx,color,comment,txfonts}
\usepackage[bookmarks=true,colorlinks,citecolor=blue,urlcolor=blue]{hyperref}
\usepackage{dsfont}

\usepackage{bbm}
\usepackage{float}

\usepackage{dcolumn}   
\usepackage{bm}        
\usepackage{filecontents}
\usepackage{lineno}

\usepackage{mathtools}
\usepackage[usenames,dvipsnames]{xcolor} 

\usepackage[export]{adjustbox}
\usepackage{adjustbox}

\usepackage{braket}
\usepackage{tikz}

\newcommand\bs[1]{\ensuremath{\boldsymbol{#1}}}

\newcommand{\eqw}[1]{(\ref{#1})}
\newcommand{\eq}[1]{Eq.\thinspace(\ref{#1})}
\newcommand{\fig}[1]{Fig.\thinspace{}\ref{#1}}
\newcommand{\fc}[1]{({#1})}
\newcommand{\figc}[2]{Fig.\thinspace{}\ref{#1}\thinspace{}\fc{#2}}

\newcommand{\plaqv}{
\tikz[baseline=-0.62ex]{
\draw [gray] (0,-0.19) -- (0.3,-0.19);
\draw [gray] (0,0.19) -- (0.3,0.19);
\filldraw [color=black, fill=black!65] (0,0) ellipse (0.06 and 0.23); 
\filldraw [color=black, fill=black!65] (0.3,0) ellipse (0.06 and 0.23); }
}

\newcommand{\plaqh}{
\tikz[baseline=-0.62ex]{
\draw [gray] (-0.19,-0.19) -- (-0.19,0.19);
\draw [gray] (0.2,-0.19) -- (0.2,0.19);
\filldraw [color=black, fill=black!65] (0,-0.15) ellipse (0.23 and 0.058); 
\filldraw [color=black, fill=black!65] (0,0.15) ellipse (0.23 and 0.058); }
}

\begin{document}

\title{Emergent Glassy Dynamics in a Quantum Dimer Model}
\author{Johannes Feldmeier}
\author{Frank Pollmann}
\author{Michael Knap}
\affiliation{Department of Physics and Institute for Advanced Study, Technical University of Munich, 85748 Garching, Germany}
\affiliation{Munich Center for Quantum Science and Technology (MCQST), Schellingstr. 4, D-80799 M{\"u}nchen, Germany}
\date{\today}

\begin{abstract}
We consider the quench dynamics of a two-dimensional quantum dimer model and determine the role of its kinematic constraints. We interpret the non-equilibrium dynamics in terms of the underlying equilibrium phase transitions consisting of a BKT-transition between a columnar ordered valence bond solid (VBS) and a valence bond liquid (VBL), as well as a first order transition between a staggered VBS and the VBL. We find that quenches from a columnar VBS are ergodic and both order parameters and spatial correlations quickly relax to their thermal equilibrium. By contrast, the staggered side of the first order transition does not display thermalization on numerically accessible timescales. Based on the model's kinematic constraints, we uncover a mechanism of relaxation that rests on emergent, highly detuned multi-defect processes in a staggered background, which gives rise to slow, glassy dynamics at low temperatures even in the thermodynamic limit.
\end{abstract}

\maketitle

\textit{Introduction.}-- Recent experimental progress on realizing and controlling synthetic quantum matter has granted new access to the rich nature of  many-body dynamics and permits the exploration of foundational aspects of non-equilibrium statistical physics~\cite{Bloch_2008_review, Polkovnikov_2011_review}. Far-from equilibrium states of strongly interacting and non-integrable quantum many-body systems are generally believed to quickly relax to a local thermal equilibrium at infinite temperature. By contrast, a recent experiment with one-dimensional ultracold atoms in the Rydberg blockade regime has found that certain highly-excited states feature long-lived coherent oscillations of local observables~\cite{Bernien_2017_51simulator}. Due to the Rydberg blockade~\cite{Urban_2009_blockade, Gaetan_2009_rydbergexcitation, Wilk_2010_rydbergentanglement, Schau_2012_rydberg2d, Schauss_2015_crystallization, Labuhn_2016_tunablerydberg, Zeiher_2017_coherentspins, Bernien_2017_51simulator}, a constrained quantum many-body system is realized in which two excitations are forbidden to occupy neighboring lattice sites. Theoretical works have proposed a set of exceptional eigenstates, entitled ``quantum many-body scars'', and nearby integrable points to be responsible for the exotic quantum dynamics~\cite{Chandran_2016_eth, Turner_2018_scars, Ho_2018_periodicorbits, Khemani_2018_rydbergintegrability, Turner_2018_scars2,Lin_2018_ETHviolation,Choi_2018_emergentSU2scars}.

\begin{figure}[t]
\centering
\includegraphics[width=.97\columnwidth]{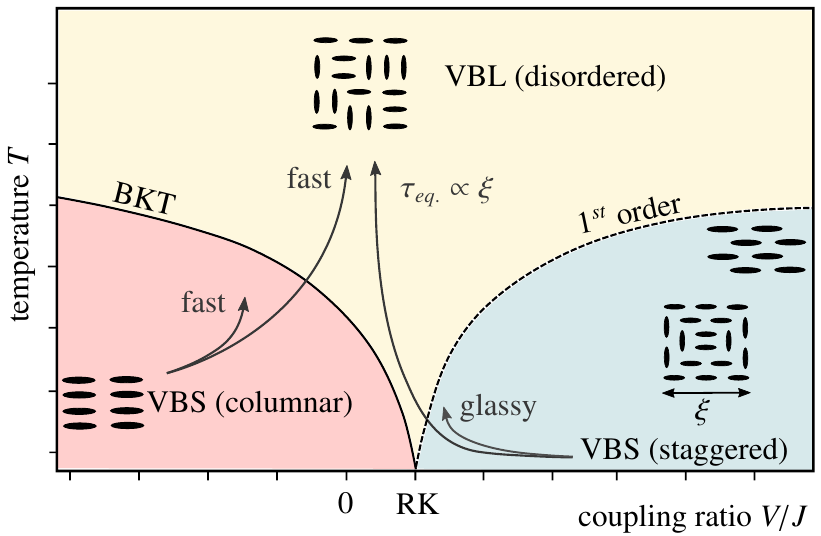}
\caption{\textbf{Dynamical phase diagram of the QDM (schematic).} Quenches within and across the columnar phase are accompanied by fast relaxation of local observables to thermal expectation values, with a rate that is fixed by the microscopic parameter $J$. Quenches across the $1^\text{st}$-order transition show a `melting' character, with thermalization times $\sim \xi$ given by the length scale of the staggered domains. Within the staggered phase, the dynamics appears glassy with relaxation times bounded by $\tau_{\mathrm{eq.}}\gtrsim\exp(c\log(V/J)\,\xi^4)$. Arrows are indicative of the different quench protocols considered.}
\label{fig:1}
\end{figure}

The 1D Rydberg chain admits a direct mapping to yet another constrained model: the close-packed dimer coverings on a two-rung ladder \cite{Chepiga_2019_DimerDMRG}. Here, we consider the far-from-equilibrium quantum dynamics of a quantum dimer model (QDM) in \emph{two spatial dimensions}; see also Refs.~\onlinecite{Lan_2018_glassy,Lan_2017_ETHdimer}. We determine the rich dynamical phase diagram upon quenching the model far from thermal equilibrium (see \fig{fig:1} for an overview) and identify initial states at finite energy density whose dynamical relaxation is obstructed as a consequence of kinematic constraints. Moreover, for such initial states consisting of a set of staggered dimer domains of length scale $\xi$, we analytically derive a lower bound on the local thermalization timescale that is exponential in $\xi^4$. Induced by highly detuned processes involving `defects' seperated by a distance $\xi$, this reveals a mechanism similar to proposals of fractonic systems at low temperatures \cite{Chamon_2005_glassiness,Nandkishore_GlassyFractons,Sala_2019_fractons}, leading to arrested quantum dynamics at $T=0$ and slow relaxation at finite energy densities reminiscent of the physics of classical glasses \cite{garrahan_2018_review}.

\textit{Model.}-- We start by introducing the QDM on the square lattice, where a hard-core constraint enforces each site to be occupied by exactly one singlet dimer, see insets in \fig{fig:1} for the illustration of a few dimer configurations. The dynamics is generated by the following Hamiltonian, originally introduced by Rokhsar and Kivelson (RK) \cite{RK_1988_dimer},
\begin{eqnarray} \label{eq:1}
\hat{H}& =& \hat{H}_V + \hat{H}_J \\
\hat{H}_V& =& V \sum_\mathrm{plaq.} \left( \Ket{\plaqv} \Bra{\plaqv} + \Ket{\plaqh} \Bra{\plaqh} \right)\nonumber \\
\hat{H}_J& =& -J \sum_\mathrm{plaq.} \left( \Ket{\plaqv} \Bra{\plaqh} + \Ket{\plaqh} \Bra{\plaqv} \right)\nonumber
\end{eqnarray}
Here, $\hat{H}_V$ gives a constant energy-offset to each pair of parallel dimers, while the off-diagonal kinetic term $\hat{H}_J$ flips a pair of resonant singlets. Importantly, the dimer model features non-local conservation laws represented by winding numbers $W_x$ and $W_y$, which provide a staggered count of the number of dimers intersecting a given straight line through the sample \cite{Moessner_2011_review}. For the remainder of this work, we restrict to the zero-winding sector $W_x=W_y=0$, which constitutes the largest part of the full Hilbert space.

The nature of the equilibrium phase diagram of the square-lattice QDM remains a matter of high interest, but seems to converge to a framework similar to the one depicted in \fig{fig:1} \cite{Leung_1996_dimerED, Alet_2005_cassicaldimer, Syljuaasen_2006_plaquette, Castelnovo_2007_BKT, Wenzel_2012_columnar, Banerjee_2014_squaredimer, Oakes_2018_dimerensembles}. In particular, for $T=0$, the model possesses an RK-point at $V=J$, where the exact ground state wave function can be constructed as an equal weight superposition of all dimer coverings within each winding number sector~\cite{RK_1988_dimer}. The RK-point separates two crystalline VBS phases, that show columnar order for $V<J$ and staggered order for $V>J$. Both VBS phases exist up to certain finite temperatures, before the transition to a disordered VBL phase, which is conjectured to extend all the way to the RK-point at $T=0$~\cite{Alet_2006_classdimer, Banerjee_2014_squaredimer}. While the columnar-VBL transition is expected to be of BKT-form, and thus continuous, the staggered-VBL transition is of $1^\text{st}$-order \cite{Alet_2005_cassicaldimer, Castelnovo_2007_BKT}.
\\

\textit{Columnar states: Thermalization.}--
The BKT-transition can be captured by introducing an order parameter that detects the spontaneous breaking of $C_4$-rotational lattice symmetry,
\begin{equation} \label{eq:2}
\hat{\phi}_c = \frac{2}{L^2}\sum_\mathrm{plaq.} \left( \Ket{\plaqv} \Bra{\plaqv} - \Ket{\plaqh} \Bra{\plaqh} \right),
\end{equation}
which counts the imbalance between horizontal and vertical plaquettes on an $L\times L$ square lattice. Restricting to the zero momentum  sector on periodic boundary conditions (PBCs), $\hat{\phi}_c$ distinguishes the two translational invariant columnar ground states $\Ket{c_A}$ and $\Ket{c_B}$, related by a $\pi /2$-rotation and $\hat{\phi}_c\ket{c_{A/B}} = \pm \ket{c_{A/B}}$.

\begin{figure}[]
\centering
\includegraphics[trim={0.cm 0cm 0.cm 0cm},clip,width=0.48\textwidth]{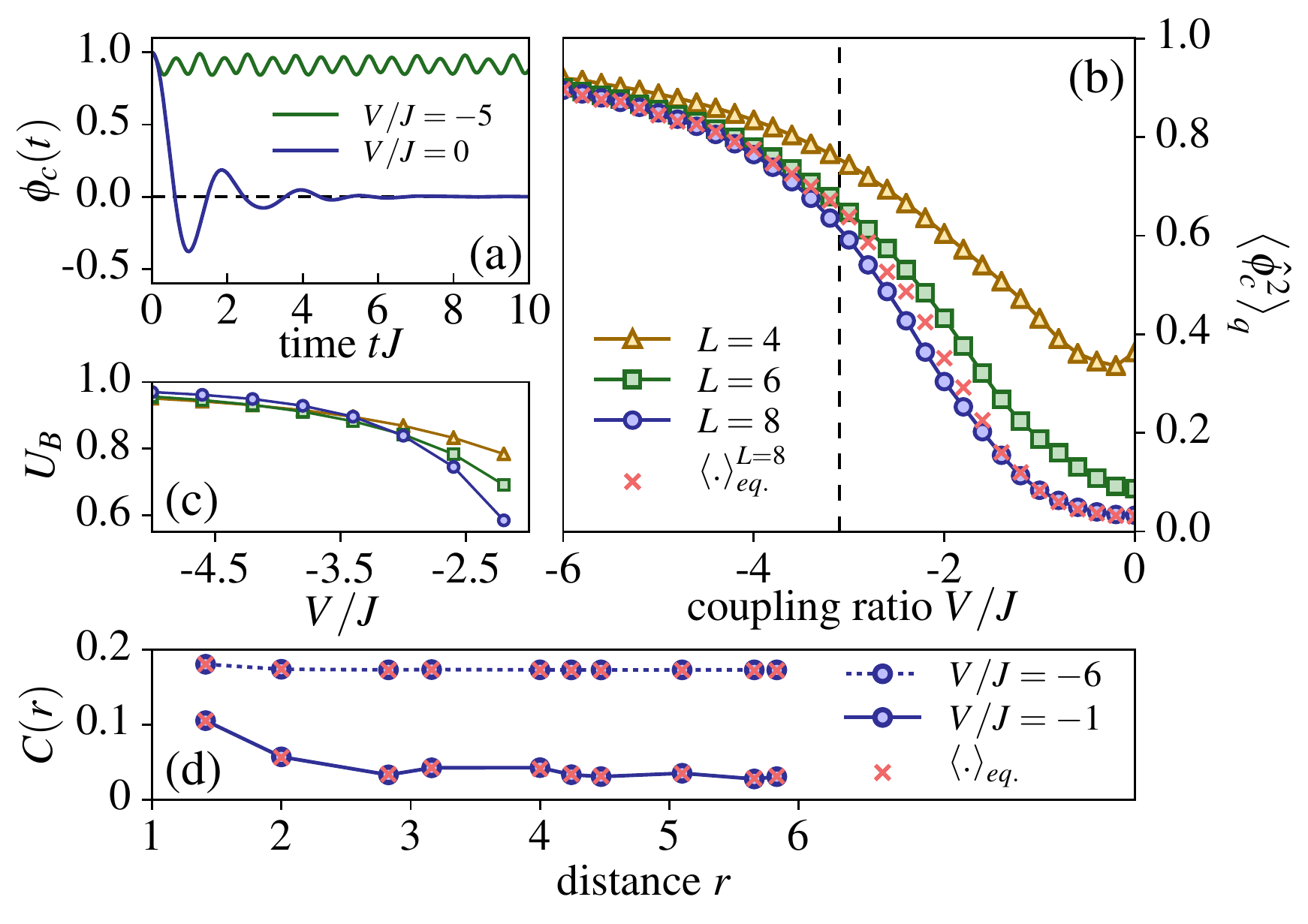}
\caption{\textbf{Thermalization of columnar initial states.} \textbf{(a)} Relaxation of the columnar order parameter $\hat{\phi}_c$ both within and across the columnar phase. \textbf{(b)} Long-time averaged values of $\hat{\phi}_{c}^2$ for different system sizes, starting from a columnar inital state and quenched to finite values of $V/J$. Included are the corresponding thermal values for $L=8$. Agreement between the two becomes less accurate around the phase transition at $V_c/J \approx -3.1$, but is excellent deep within the phases. \textbf{(c)} The crossing in the long-time averaged Binder cumulant $U_B$ of $\hat{\phi}_{c}$ marks the corresponding finite-$T$ phase transition. \textbf{(d)} The time-averaged dimer correlation functions $C(r)$, starting from a columnar initial state, agree well with the corresponding equilibrium values, even deep within the VBS phase. }
\label{fig:2}
\end{figure}

\begin{figure}[t]
\centering
\includegraphics[width=.95\columnwidth]{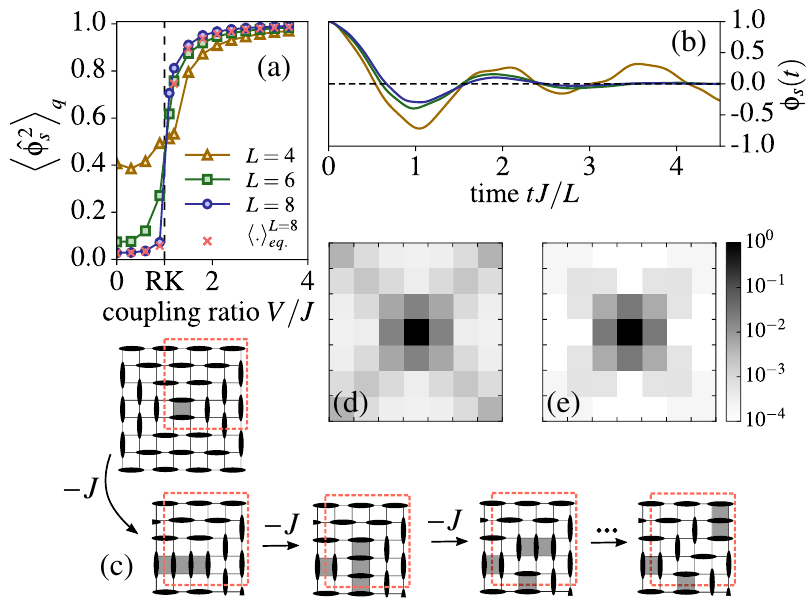}
\caption{\textbf{Localization of staggered initial states.} \textbf{(a)} Long-time averaged values of the staggered order parameter $\hat{\phi}_{s}^2$ starting from a staggered initial state. \textbf{(b)} In a quench across the transition to the disordered regime (here: $V/J=0$), the dynamics of $\hat{\phi}_{s}$ scales with $L^{-1}$.
\textbf{(c)} The state $\ket{p_A}$. Excitations from this state containing four plaquettes are expected to delocalize along the diagonals of the system. \textbf{(d)} The thermal expectation value of the potential energy landscape at an energy matching the quench in \fc{e}. \textbf{(e)} The late-time plaquette densities obtained after quenching $\frac{1}{\sqrt{2}}(\ket{p_A}+\ket{p_B})$ to $V/J=3$ differ strongly from the thermal values in \fc{d} at large distances from the center.}
\label{fig:3}
\end{figure}

To study thermalization in our system, we compute the relaxation of order parameters in a quench protocol to their corresponding thermal expectation values. We thus consider the long-time averaged values
\begin{equation} \label{eq:3}
\Braket{\hat{O}}_q := \lim_{t\rightarrow \infty} \frac{1}{t} \int_0^tdt'\Bra{\psi(t')}\hat{O}\Ket{\psi(t')}
\end{equation}
of a given observable $\hat{O}$ following a quench from an initial state $\Ket{\psi(t=0)}=\Ket{\psi_0}$ on systems of linear size $L\in\{4,6,8\}$, which we choose as a columnar state on PBCs, $\ket{\psi_0}=\ket{c_A}$.
The averages of \eq{eq:3} are to be compared with the corresponding thermal expectation values
\begin{equation} \label{eq:4}
\Braket{\hat{O}}_\beta = \text{Tr}\left\{e^{-\beta\hat{H}}\hat{O}\right\}, \quad \Braket{\hat{H}}_\beta \stackrel{!}{=} \Bra{\psi_0}\hat{H}\Ket{\psi_0},
\end{equation}
where the effective inverse temperature $\beta$ is chosen to match the energy of the initial state. For $L=8$, the system cannot be diagonalized fully, so we use the typicality approach~\cite{Bartsch_2009_typicality} where the expectation values of \eq{eq:4} are drawn from random (infinite temperature) initial states, which are subsequently evolved in imaginary time up to $\tau=\beta$ such that \eq{eq:4} is fulfilled. We comment on the effective temperatures obtained from \eq{eq:4} in the supplementary material \cite{supplementary}.

As displayed in \figc{fig:2}{a+b}, the columnar order parameter shows efficient relaxation to thermal values, less accurate only in the vicinity of the arising phase transition. 
The thermalization of order parameters allows for the determination of finite-temperature phase transitions via finite size scaling arguments, as has been done for ground state phases of the model \cite{Leung_1996_dimerED}. Even though the exact transition point between columnar and disordered phase may turn out quite inaccurate in such small systems \cite{Leung_1996_dimerED, Banerjee_2014_squaredimer}, the qualitative picture is expected to hold, nonetheless. As shown in \figc{fig:2}{c}, the Binder cumulant $U_B \propto 1-\Braket{\hat{\phi}_c^4}_q/3\Braket{\hat{\phi}_c^2}_q^2$ shows a crossing at $V/J\approx -3.1$, which signals the transition to the $C_4$-symmetry-breaking columnar phase.
Time-averages were converged at times $tJ=300$ on $L=8$, which supports a quick thermalization of $\hat{\phi}_c$ in a wide range of model parameters. 

This fast thermal relaxation can be substantiated by investigating the behaviour of local observables. For this purpose, we define the dimer-dimer correlation functions $C(r)$ as
\begin{equation} \label{eq:5}
C(r) = \Braket{\hat{n}_0\hat{n}_{\bs{r}}} - \Braket{\hat{n}_0}\Braket{\hat{n}_{\bs{r}}},
\end{equation}
where $\hat{n}_{\bs{r}}$ is the dimer occupation number at bond $\bs{r}$. Quenches from $\ket{c_A}$ even deep inside the columnar phase are accompanied by fast relaxation of $C(r)$, save for spontaneous symmetry breaking. The latter can be accounted for by choosing a rotationally invariant initial state $\ket{\psi_0}=\frac{1}{\sqrt{2}}(\ket{c_A}+\ket{c_B})$, which leads to the results of \figc{fig:2}{d}. 

The fast thermalization results from the fact that the columnar initial state is the configuration with the largest number of flippable plaquettes, and hence the kinetic part $\hat H_J$ of the Hamiltonian \eqw{eq:1} is able to effectively explore the phase space.
\\

\textit{Staggered states.}--
Quite to the contrary, the fully staggered state is part of a maximum-winding sector and entirely frozen. We thus construct the exact groundstate in the limit $V/J\rightarrow \infty$ within the zero-winding sector, which yields a state of `pyramid'-like shape, where the tip of the pyramid serves as a dynamically active defect between extended areas of staggered configurations, see \fig{fig:1} inset. 
In the thermodynamic limit on PBCs, the staggered ground state contains four such pyramidic domain walls \cite{Oakes_2018_dimerensembles}. For a quench from the staggered phase to the disordered regime however, the dynamics will mainly be governed by the behavior of individual pyramid states, which we can effectively capture by considering a single pyramid on open boundary conditions (OBCs). This choice allows us to double the length scale $\xi$ of the initial state, given our finite size limitations. On an OBC-geometry, there exist two pyramidic ground states $\Ket{p_A}$ and $\Ket{p_B}$, from which we construct a $\mathbb{Z}_2$ order parameter to distinguish them:
\begin{equation} \label{eq:6}
\hat{\phi}_s:=\frac{2}{L^2}\left\{\sum_{l_A}\hat{n}_{l_A}-\sum_{l_B}\hat{n}_{l_B}\right\}.
\end{equation}
Here, the indices $l_A$ denote the bonds occupied in the $\Ket{p_A}$-state, $\hat{n}_{l_A}\Ket{p_A} = \Ket{p_A}$, and correspondingly for $l_B$, such that again $\hat{\phi}_s\ket{p_{A/B}} = \pm \ket{p_{A/B}}$. In a given quench protocol starting from e.g. $\Ket{p_A}$, the computation of $\Braket{\hat{\phi}_s(t)}$ thus involves the memory-function of the initial state, and (non-)relaxation of the same is intuitively connected to spontaneous symmetry breaking. Even though the Hilbert space is somewhat restricted on OBCs, we still expect the dynamics of the $1^\text{st}$-order transition to be well captured in our approach, as the dynamics is initiated in the center of the pyramid. 

Our quench protocol crosses a phase transition at an infinitesimal distance from the RK point at $V=J$, see \figc{fig:3}{a}, due to the vanishing energy density of the pyramidic initial state; see our supplementary material \cite{supplementary}. Therefore, the dynamics of $\hat{\phi}_s$ signals a $1^\text{st}$-order transition at the RK-point, with a sharp rise of $\Braket{\hat{\phi}_s^2}_q$ for larger systems. 
Quenching across the transition to the disordered phase, the dynamical order parameter in \figc{fig:3}{b} shows an approximate collapse of its zero-crossings and extrema upon rescaling time $t\rightarrow t/L$. This characterizes the melting dynamics proliferating around the dynamical center of the initial state, which successively has to work its way to the outside, and implies a timescale $\tau_{eq.}$ of thermalization that is effectively set by the length scale $\xi$ of the initial staggered domain. This scale also impacts the Loschmidt return rate that can be used to characterize the dynamical phase transitions, see supplementary material \cite{supplementary}.

\textit{Localization.}-- While monitoring the order parameter suggests thermal behavior also within the staggered phase, \figc{fig:3}{a}, this is not the case for local observables which indicate the absence of thermalization for such quenches. This can be demonstrated by mapping out the potential energy landscape, i.e., the plaquette density, following a quench. Here, we start from a symmetrized staggered state $\ket{\psi_0}=\frac{1}{\sqrt{2}}(\ket{p_A}+\ket{p_B})$, quenching to $V/J=3$. We find the thermal potential energy landscape to include delocalized plaquettes along the system diagonals, \figc{fig:3}{d}. In sharp contrast, the distribution of plaquettes following the quench, averaged up to $tJ=300$ on $L=8$, remains localized around the center, see \figc{fig:3}{e}. An analysis on $L=6$, where the system may be diagonalized fully, also shows no signs of thermalization at later times. 

To understand this property, we can consider the large-$V/J$ dynamics around, say, $\ket{p_A}$ as an effective single-particle problem on a finite, 1D lattice in a potential $U_i$, where $U_0=V$, $U_1=3V$, and $U_{i\geq2}=4V$ counts the plaquettes. Here, $i=0$ corresponds to $\ket{p_A}$, $i=1$ labels the second state of \figc{fig:3}{c} with three flippable plaquettes, and all states with $i\geq 2$ have four plaquettes. For sufficiently strong $V/J$, the large overlap of the initial state $\ket{\psi (0)}=\ket{i=0}$ with the bound states of the effective potential well $U_i$ located around $i=0$ defies global thermalization and leads to an enhanced localization of the long-time averaged state $\ket{\psi(t)}$, in good agreement with the results of \fig{fig:3}. More details are provided in the supplementary material~\cite{supplementary}. We have observed similar effects in the dynamics of columnar states containing string-like excitations, composed of finite columns aligned perpendicular to the background state. These states are strongly repulsively bound, leading to large relaxation time scales. 
In either case, constraints are essential for the emergent slow quantum dynamics.

\textit{Thermodynamic limit.}-- A natural question arising in the context of localization is its stability in the thermodynamic limit, and whether the apparent localization on our finite systems has to be interpreted rather as a long-lived, prethermal plateau for $L\rightarrow \infty$. In this regard, we distinguish two ways to take the thermodynamic limit: (A) We take $L,\xi\rightarrow \infty$ such that $L/\xi\rightarrow \mathrm{const.}$, which corresponds to a finite number of pyramids in the initial state and hence a vanishing energy density for $V/J>1$, see \cite{supplementary}. (B) We let $\xi= \mathrm{const.}$ such that $L/\xi\rightarrow\infty$, which leaves us with a finite density of pyramids and thus a finite energy density in the thermodynamic limit.

Case (A) is captured by the numerical results for small systems, where the appearance of bound states within each pyramid ensures the localization around its respective center. The independence of the pyramids with respect to each other is ensured by their diverging lengthscale $\xi$, effectively protecting them against melting. This last point can be made more specific by arguing that the matrix elements between certain states with different $\xi$ in a given energy shell are highly suppressed compared to their level spacing, similar in spirit to arguments employed in the context of MBL \cite{DeRoeck_2017_mblRG}.
In particular, there exist states that are close in energy, yet not exactly degenerate, which can only be connected via $\mathcal{O}(L^2)$ consecutive plaquette flips that are off-resonant from the starting energy shell. Such processes correspond to reorderings of the initial state (see supplementary material \cite{supplementary}). The resulting matrix elements $\Delta\sim J(J/V)^{\mathcal{O}(L^2)}$ can thus always be tuned smaller than the relevant many-body level splitting $\epsilon\gtrsim J\exp(-const.\times L^2)$ for $L\rightarrow \infty$, leading to non-hybridized states and thus, arrested quantum dynamics at $T=0$.

\begin{figure}[t]
\centering
\includegraphics[width=.99\columnwidth]{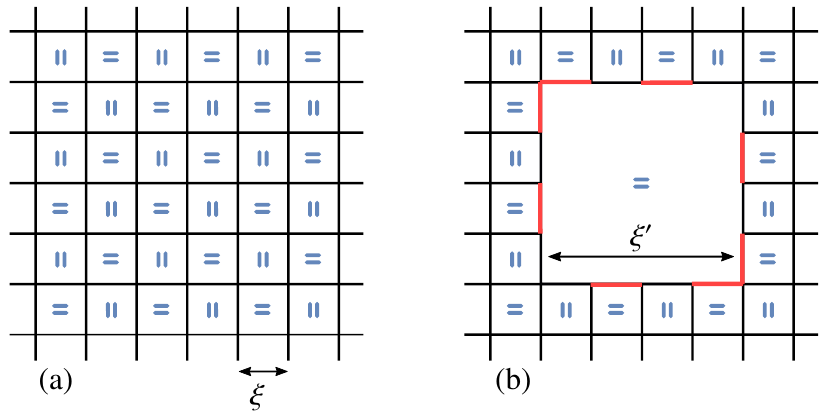}
\caption{\textbf{Cost for local equilibration.} \textbf{(a)} Starting configuration of pyramids of length $\xi$, where the orientation of the central plaquettes is indicated in blue. \textbf{(b)} Thermalization of local observables requires the formation of larger pyramids of size $\xi'$, which creates defects (red) with respect to the original background at the boundaries.}
\label{fig:4}
\end{figure}

In case (B), taking place at finite energy densities, we can provide a simple argument that yields a lower bound on the thermalization timescales of \emph{local} observables.
To this end, we start from an initial state with a small, but finite energy density $\varepsilon \sim 1/\xi^2$, distributed equally over pyramids of fixed length $\xi$, see \figc{fig:4}{a}. The thermalization time $\tau_{\mathrm{eq.}}$ is then given by the largest matrix element $\Delta$ that hybridizes states with equal energies but locally different $\xi$, which in turn is bounded from above by the lowest order in perturbation theory at which such similar energy states can be reached.

Given the initial state of homogeneous energy density, we ask at which timescale local inhomogeneities can arise, leading to relaxation of local observables. In terms of the staggered dimer inital state, this time corresponds to the time needed in order to break up the pyramids. To keep the overall energy balanced, all processes that lead to states of similar energy necessarily need to break up pyramids of size $\xi$ and reassemble them as pyramids of larger size $\xi'$, thereby reducing the local energy density. As shown in \figc{fig:4}{b}, this will create a number of defects along the boundary of this new formed pyramid in the background of the original staggered lengthscale $\xi$. The total energy difference of this process can then be estimated by the difference in the number of parallel dimer plaquettes,
$\Delta E/V \simeq \left\{1-\left(\frac{\xi'}{\xi}\right)^2\right\}+\left[\frac{\xi'}{\xi}\left(\xi-1\right)\right]$,
where the term in the curly brackets originates from the reduced number of plaquettes within the larger pyramid, while the term in the square brackets is caused by the defects along the new boundary.
To reach states of similar energy, we set $\Delta E = 0$ and obtain
$\frac{\xi'}{\xi} = \xi+\mathcal{O}(1/\xi)$.
The minimum size of the new pyramid is therefore given by $\xi'\sim \xi^2$. Due to the staggered nature of the system, this state can only be reached via un- and refolding the area of the new pyramid, which requires $\sim \xi'^2$ plaquette flips, all off-resonant from the starting energy. The matrix element can thus be shown to be at most proportional to
\begin{equation} \label{eq:7}
\Delta \sim J\left(\frac{J}{V}\right)^{c\,\xi'^2} = J \exp\left\{-c\log\left(\frac{V}{J}\right)\,\xi^4\right\},
\end{equation}
where $c$ is a constant of order $\mathcal{O}(1)$. We observe that the implied lower bound for the time scale $\tau_{\mathrm{eq.}}\sim 1/\Delta$ grows extremely fast with the staggered scale $\xi$ of the inital state for $V/J>1$. The system sizes required to observe the slow relaxation of \eq{eq:7} at large $V/J$ are numerically out of reach. Accordingly, small scale simulations of systems up to L=8 sites with pyramids of size $\xi \sim 4$ show the absence of relaxation, as predicted by our analytical formula and demonstrated numerically above as well as in our supplementary \cite{supplementary}. \eq{eq:7} should be regarded as a lower bound for the relaxation time, which is why formally we cannot exclude full localization.
\\

\textit{Conclusion \& Outlook.}-- In this work, we have examined the dynamical phase diagram of the square lattice quantum dimer model.
We have found that inside the staggered phase, local relaxation is lower bounded by an extraordinarily large time scale characteristic for glassy systems, valid even in the thermodynamic limit for quench dynamics initiated even with a finite energy-density initial state. The associated mechanism employs emergent non-dispersing defects at low energies, possibly signaling connections to the physics of fractons to be studied in future works.

Future lines of investigation may also address the existence of quantum-many body scars in two dimensions. In particular, determining the dimensional crossover from the two-leg ladder to the fully two-dimensional quantum dimer model may provide further insights. 
Interesting questions on the dynamical nature of dimerized systems also naturally extend to the various manifestations of the quantum dimer model on different lattice geometries as well as different winding sectors. In particular, the presence of local constraints permits to study various types of emergent dynamical gauge fields besides the $U(1)$-liquid of the square lattice. Direct connections between constrained models in dimensions larger than one and the controllable dynamics of Rydberg-blockaded atoms would thus be much desirable, as they could possibly open up new roads to the experimental investigation of fundamental matters in nonequilibrium quantum physics.\\

\textit{Acknowledgments.}--
We thank Claudio Castelnovo, Ignacio Cirac, Juan Garrahan, Sarang Gopalakrishnan, Clemens Kuhlenkamp, Andreas L\"auchli and Nicola Pancotti for insightful discussions. 
We acknowledge support from the Technical University of Munich - Institute for Advanced Study, funded by the German Excellence Initiative and the European Union FP7 under grant agreement 291763 (M.K.), the Deutsche Forschungsgemeinschaft (DFG, German
Research Foundation) under Germany's Excellence Strategy -- EXC-2111-390814868 (F.P., M.K.), from the DFG grant No. KN 1254/1-1 (J.F., M.K.), the DFG TRR80, Project F8 (J.F., F.P., M.K.), the DFG Research Unit FOR 1807 through grant no. PO 1370/2- 1 (F.P.), the Nanosystems Initiative Munich (NIM) by the German Excellence Initiative (F.P.), and the European Research Council (ERC) under the European Unions Horizon 2020 research and innovation program grant agreement no. 771537 (F.P.).

\bibliography{DDPT}


\onecolumngrid
\begin{center}
\line(1,0){250}
\section{\normalsize Supplementary Material}
\end{center}
\twocolumngrid

\subsection{\normalsize Effective Temperature} \label{localization}
To gain a better understanding of the quenches indicated in the schematic dynamical phase diagram of \fig{fig:1}, we investigate the effective temperatures as obtained from \eq{eq:4} of the main text. 
\\ \\
In the limit of $V/J \rightarrow \mp \infty$, both the columnar and staggered initial states yield $T=0$, as each of them correspond to the ground states in the respective case. By contrast, at $V/J=0$, $\Braket{\mathrm{col/pyr}|\hat{H}_{V/J=0}|\mathrm{col/pyr}}=0$ for both states. Since the spectrum of $\hat{H}$ is symmetric at this point, both states (just as every other product state) correspond to infinite temperature. Varying $V/J$ in between can give rise to finite $T$.

While for the columnar initial state, the temperature will rise strictly monotonously from $0$ to $\infty$ when lowering $|V|/J$, the same is not true for the staggered ground state. In particular, we show that for $V/J>0$, the energy of a given state $\ket{\psi}$ can be written as
\begin{equation} \label{eq:sm1}
\braket{\hat{H}}_{\psi} = \braket{\hat{H}_{RK}}_{\psi} + (V-J)\braket{\hat{H}_V}_{\psi} \geq (V-J)\braket{\hat{H}_V}_{\psi},
\end{equation}
where $\hat{H}_{RK}$ denotes the Hamiltonian at the RK-point $V/J=1$, and we have used that $\hat{H}_{RK}$ is positive definite as it can be written as a sum of projectors. From \eq{eq:sm1} we infer that for all $V/J>1$, the system's ground state cannot have an extensive amount of flippable plaquettes. This follows from the energy of such a state being extensive, while the pyramid state only has finite energy $V$ for all system sizes and ratios $V/J$. The energy difference between the exact ground state at a general coupling ratio within the staggered phase and the maximally staggered product initial state is thus only intensive, which means that in the quench protocol to finite $V/J>1$, the effective temperature is only lifted infinitesimally.

\begin{figure}[t]
\centering
\includegraphics[width=.85\columnwidth]{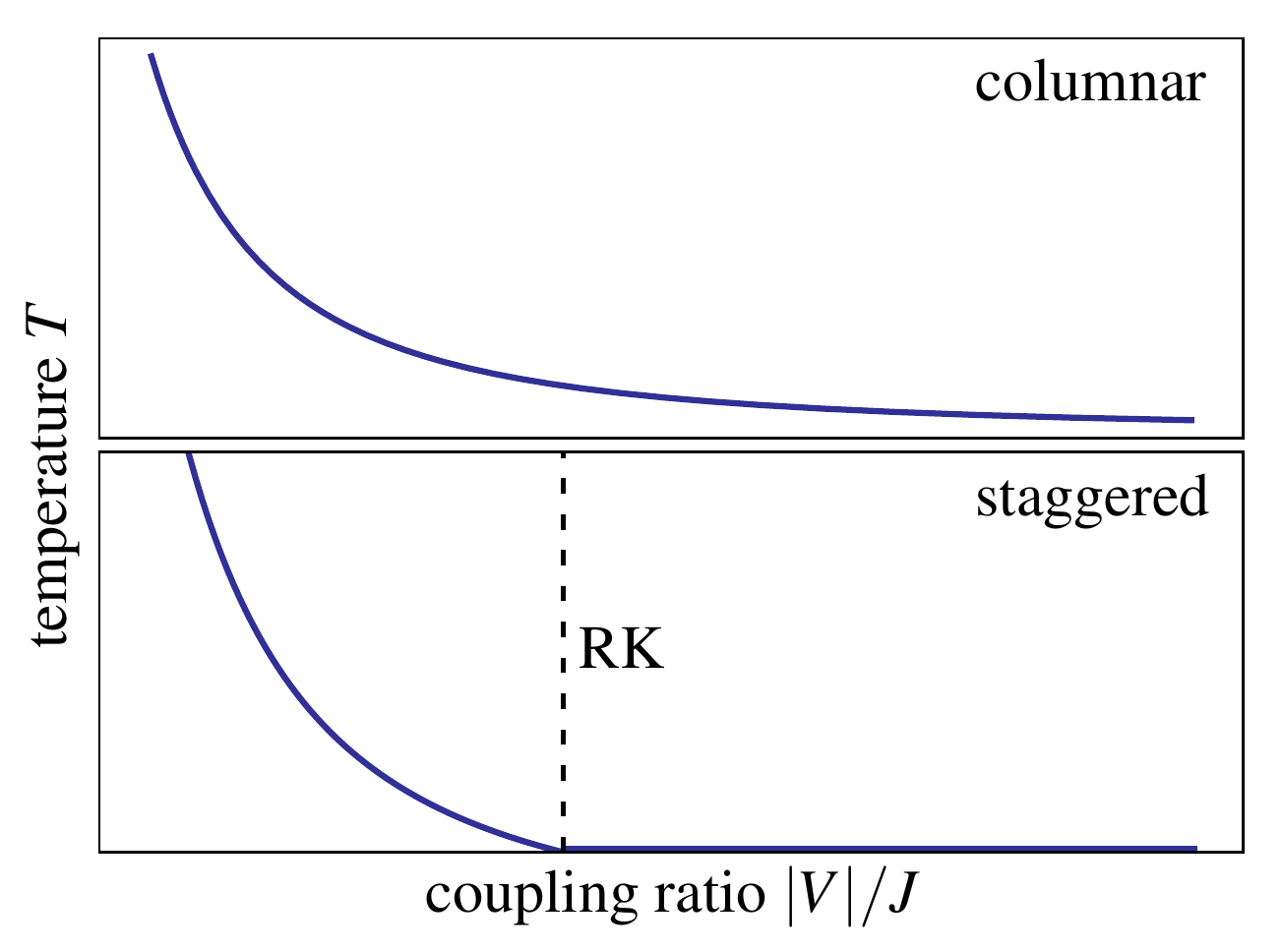}
\caption{\textbf{Effective temperatures following a quantum quench.} Schematic of the effective temperature in the thermodynamic limit of the columnar and pyramidic initial states at the corresponding quench parameters. Along $V/J>1$, the staggered initial state carries only an intensive amount of energy with respect to the ground state, and thus corresponds to an infinitesimal temperature.}
\label{fig:5}
\end{figure}

At the RK-point, the ground state is the equal weight superposition of all dimer coverings in the zero-winding sector and contains an extensive amount of plaquettes. Thus, upon crossing the RK point, i.e., for $V/J<1$, the ground state energy is bounded from above by
\begin{equation} \label{eq:sm2}
E_{\mathrm{gs}}(V/J<1) \leq -|V-J|\braket{\hat{H}_V}_{\psi = \mathrm{RK}},
\end{equation}
and is thus extensive. Hence, once the RK-point is crossed in the quench, the amount of energy inserted into the system by the quench is extensive, and  corresponds to finite temperatures. This shows directly that the transition at the RK-point to the staggered phase is also discontinuous within the zero-winding sector.

Beyond these considerations, the effective temperature can in principle be extracted numerically from \eq{eq:4} of the main text. A schematic of the effective temperature is shown in \fig{fig:5}.

As a final remark for this section, we stress that although quenches inside the staggered phase starting from the pyramidic ground state add only infinitesimally to the temperature in the thermodynamic limit, quenches from an initial state with finite staggered scale $\xi$ do add an extensive amount of energy to the system and correspond to finite temperatures. Therefore, the non-thermal dynamical properties of the quantum dimer model found numerically in the main text correspond to a dynamical arrest at $T=0$, characteristic of structural glasses.

\subsection{\normalsize Dynamical Phase Transitions} 
\begin{figure*}[t]
	\centering
	\includegraphics[trim={1.2cm 0.0cm 1.4cm 0.8cm},clip,width=.99\textwidth]{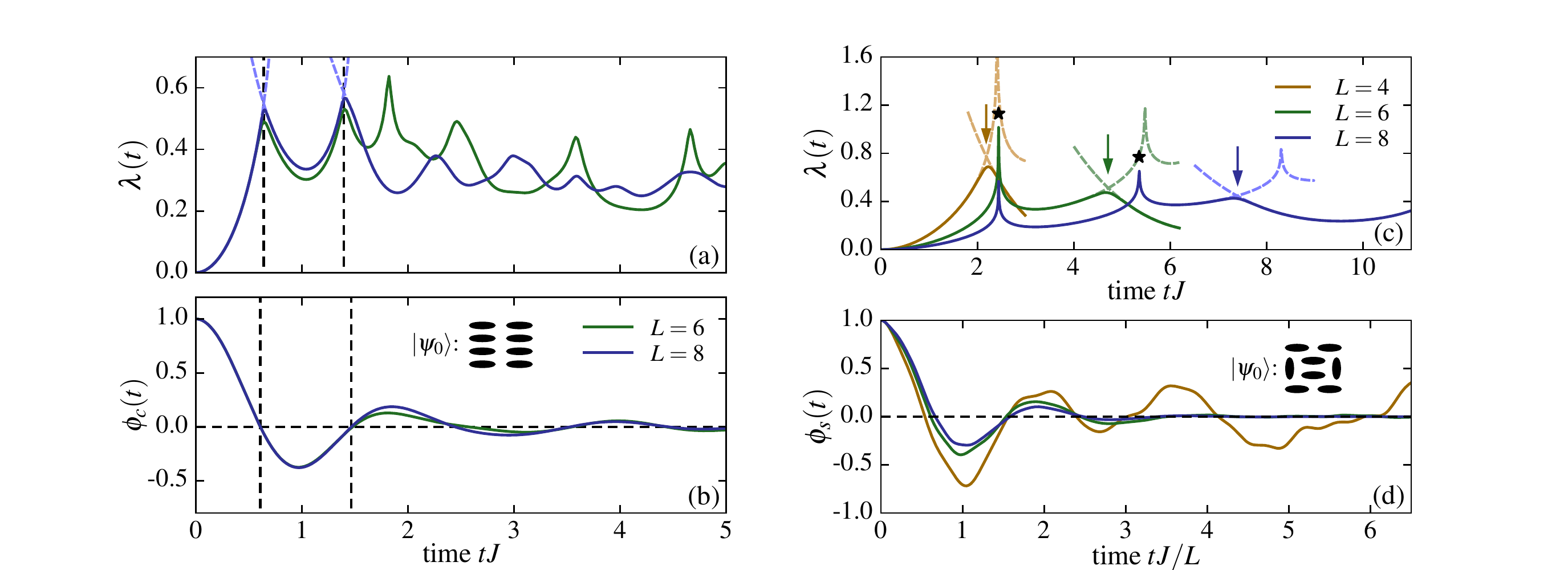}
	\caption{\textbf{Dynamical Phase Transitions.} \textbf{(a)} Loschmidt rate $\lambda(t)$ for $L=6,8$ following a quench from of a columnar product state to $V=0$. The light blue dashed lines indicate the weights $\lambda_{A/B}$ of the individual ground states $\ket{c_{A/B}}$. \textbf{(b)} Columnar order parameter dynamics for the same quench. The dashed lines mark the zero-crossings of $\phi_c(t)$ and kinks of $\lambda(t)$, respectively. \textbf{(c)} shows $\lambda(t)$ for $L=4,6,8$ starting from a staggered initial state, with light dashed lines corresponding to the weights $\lambda_{A/B}(t)$ of $\ket{p_{A/B}}$ in the vicinity of their crossing points, marked by arrows. Black stars mark resonances. \textbf{(d)} shows $t/L$-scaling of the staggered order parameter dynamics.}
	\label{fig:6}
\end{figure*}
As discussed shortly in the main text, we can use the theory of dynamical phase transitions (DPTs)~\cite{Heyl_2013_transverseIsing} to further characterize and compare the BKT- and the $1^{st}$-order transitions. A hallmark in the study of DPTs is the emergence of non-analytic, kink-like structures in the return amplitude to the original ground state manifold following a quench, the Loschmidt rate, defined as
\begin{equation} \label{eq:sm3}
\lambda(t)=-\frac{2}{L^2}\log\left(\sum_{n\in \{gs\}}|\Braket{n|\psi(t)}|^2\right).
\end{equation}
Here, the manifold $\{gs\}$ consists of $\{\ket{c_A},\ket{c_B}\}$ for the BKT, and $\{\ket{p_A},\ket{p_B}\}$ for the $1^\text{st}$-order transition. The behavior of $\lambda(t)$ upon quenching across the BKT-transition is shown in \figc{fig:6}{a}, where an initial columnar state is taken to the VBL phase at $V=0$. The time evolution of $\phi_c(t)$ has already converged reasonably well for $L=6,8$, while $\lambda(t)$ exhibits significant finite size fluctuations which dominate the Loschmidt rate at late times. This is evidenced by the suppressed magnitude of oscillations at late times on $L=8$ as compared to $L=6$. Nonetheless, there exist two systematic crossings of the individual ground state weights $\lambda_{A/B}(t)=-2/L^2\log|\left<c_{A/B}|\psi(t)\right>|^2$, at which sharp kinks are expected to form for $L\rightarrow\infty$. The critical times of the kinks are in rough agreement with the zeros of the order parameter $\phi_c(t)$, \figc{fig:6}{b}, linking the dynamical phase transitions in the order parameter and the Loschmidt echo~\cite{Zunkovic_2018_dpt, PhysRevB.96.134313}.

An analogous analysis carried out for the $1^{st}$-order DPT yields vastly different results, shown in \figc{fig:6}{c}, where the rate $\lambda(t)$ is given for $L=4,6,8$, in a quench of $\ket{p_A}$ to $V=0$. Here, we can identify the following:
First, the sharp features visible in $\lambda(t)$, marked by black stars in \figc{fig:6}{c}, correspond to resonances specific to the point $V=0$. They can be understood in a simplified picture of the pyramid tip as a single plaquette embedded in a staggered, and thus anisotropic, mean field background. 
The dynamics of the central plaquette is then described by an effective two-level Hamiltonian $H_{\mathrm{eff}}\propto -J\sigma_x+ V\sigma_z$, which, for $V\neq 0$, sustains a finite population of the inital ground state at any time. For $V=0$, the background becomes isotropic and the central plaquette undergoes coherent Rabi-oscillations, which depopulate the ground state with period $\pi/J$, roughly corresponding to the separation of the black stars in \figc{fig:6}{c}. The temporal positions of the resonances are similar on all system sizes $L$, which indicates that the short-time dynamics following the quench is dominated by only small regions around the central plaquette. This again characterizes the melting dynamics proliferating around the dynamical center of the initial state, which successively has to work its way to the outside. 

The scaling of this process is revealed by the dynamical order parameter $\phi_s(t)$ in \figc{fig:6}{d}, which shows an approximate collapse of the zero-crossings and extrema of $\phi_s(t)$ upon rescaling $t\rightarrow t/L$. Therefore, considering an initial state not composed of a single pyramid covering the lattice, but rather multiple pyramids of average size $\xi$, the timescale $\tau_{eq.}$ of thermalization is effectively set by $\xi$. Relating to the phase diagram of Fig.~1, this corresponds to a quench across the transition not from the ground state at $V/J\rightarrow\infty$, but an initial state with energy corresponding to a finite-$T$ state at a finite ratio $V/J$.

Finally, letting $L\rightarrow\infty$ and $\xi \to \infty$, the $\propto L$-scaling of the relaxation dynamics also shifts the crossing of $\lambda_{A/B}(t)$ to increasingly late times. Hence, while $L\rightarrow\infty$ can be shown to yield $\lambda(t)=\text{min}(\lambda_A(t),\lambda_B(t))$, thus developing sharp features, the angle at which $\lambda_{A/B}(t)$ cross becomes increasingly flat. If, based on the scaling of the order parameter, we conjecture a corresponding $\propto 1/L$-scaling of this angle, a simple scaling analysis for $\lambda(t)$ shows the cancellation of both effects and no discontinuity develops in $\partial_t\lambda(t)$ for $\xi\sim L\to\infty$. \\

\subsection{\normalsize Frozen Dynamics in the Staggered Phase}
In this section we provide supplementing information regarding the dynamics of the pyramidic initial state (or more general: initial states with vanishing energy density). In particular, we present details for the formation of bound states and discuss how this mechanism can lead to ergodicity breaking.
\subsection{Effective 1D system}

\begin{figure}[t]
\centering
\includegraphics[width=.97\columnwidth]{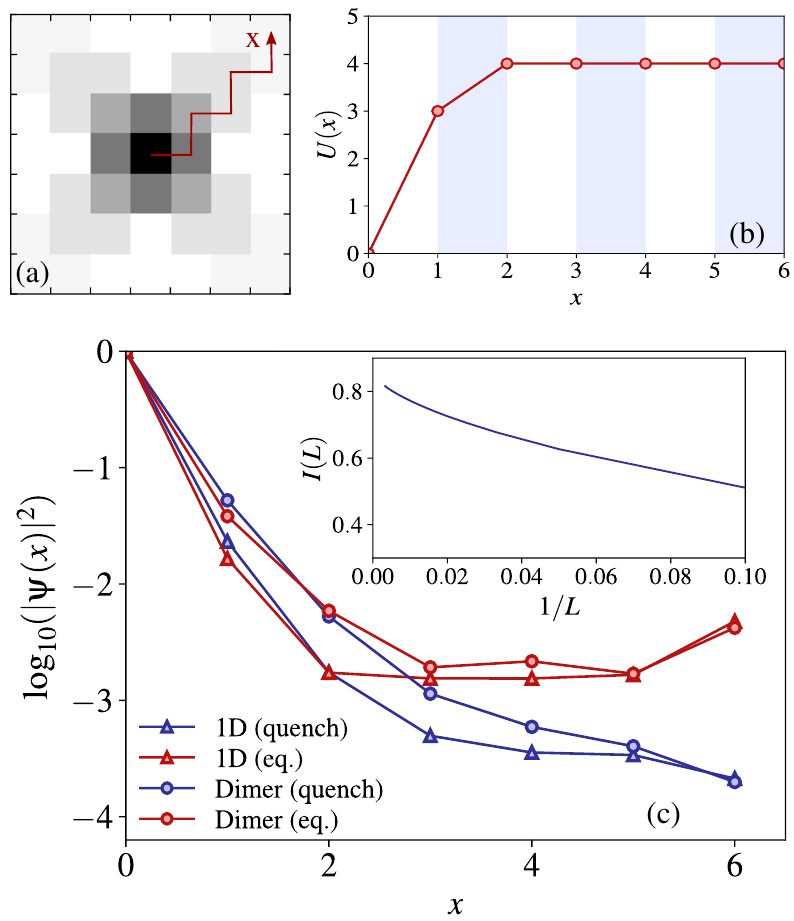}
\caption{\textbf{Effective 1D system.} \textbf{(a)} Shifting plaquettes along the diagonal of the square lattice can be interpreted as an effective 1D model, with a potential well located around the origin \textbf{(b)}. 
\textbf{(c)} Comparison of the resulting plaquette densities of the dimer model along the diagonal and its truly 1D counterpart for $V/J=3$. In both cases the equilibrium configurations possess a larger uniform weight for large enough distances compared to the quenched states. Inset: The imbalance $I(L)$ of localized plaquette-weight between quenches and equilibrium goes to a finite value in the thermodynamic limit in the 1D system.}
\label{fig:7}
\end{figure}

We consider quenches from the $\frac{1}{\sqrt{2}}(\ket{\mathrm{p_A}}+\ket{\mathrm{p_B}})$ initial state inside the staggered phase ($V/J>1$). As can be seen by \eq{eq:sm1}, the ``unfolding''  of the staggered order comes at an increasing cost of potential energy. Thus, for large $V/J$, we expect the pyramid structure to be stable in the quench. However, as discussed in \fig{fig:3} of the main text, the system can rearrange in such a way that plaquettes shift along the diagonals of the system without destroying the staggered order, thus keeping the potential energy constant. One might therefore expects delocalization along the diagonals of the system, which is indeed found in the associated thermal expectation values of \figc{fig:3}{d}. 

However, in the quench dynamics we find an apparent localization at late times. In order to understand this behavior we develop an effective one-dimensional single-particle model. Shifting the plaquettes along the 2D diagonal then corresponds to a quantum particle moving in 1D, where the position $x$ of the particle relates to the position of plaquettes. Hence, we identify $\ket{x=0} \sim \frac{1}{\sqrt{2}}(\ket{\mathrm{p_A}}+\ket{\mathrm{p_B}})$ with the pyramid initial state, while $\ket{x\geq 1}$ correspond to the states obtained by repeatedly shifting plaquettes along the diagonal, indicated in \figc{fig:3}{c} of the main text. The effective 1D potential $U(x)$ is then given by
\begin{equation} \label{eq:sm4}
U(x) =
\begin{cases}
V, \quad\;\; x=0\\
3V, \quad x=1\\
4V, \quad x\geq 1\\
\end{cases},
\end{equation}
which will host at least one bound state for finite $V$. Comparing this simple one-dimensional problem to the full dimer dynamics we find good agreement, see \figc{fig:7}{c}. In particular, in both cases the (formal) equilibrium expectation values remain constant at a certain distance from the center, whereas the results obtained from the quench dynamics continue decreasing with distance $x$. This can be attributed to the formation of bound states appearing in the potential well \figc{fig:7}{b} for sufficiently large values of $V$.

Furthermore, we can track the behaviour of the effective 1D description for large system sizes. We first note that because of $V/J>1$, for both late time quench and equilibrium, most of the plaquette-weight still resides in the center of the pyramid. We hence assume the plaquette-density $f^L_\tau(x)$ at system size $L$ for both quench ($\tau=\mathrm{q.}$) and equilibrium ($\tau=\mathrm{eq.}$) to be the sum of a dominant exponentially localized part as well as a delocalized tail that originates from the scattering states of $U(x)$,
\begin{equation} \label{eq:sm5}
f^L_{\tau}(x) = a^L_\tau e^{-b^L_\tau\cdot x} + \frac{c^L_\tau}{L}, \quad \tau \in \mathrm{\{q.,eq.\}},
\end{equation}
with $a^L_\tau,b^L_\tau,c^L_\tau$ system size dependent constants. Then, the small amount of additional plaquettes compared to the single plaquette of the initial state, generated by either quench dynamics or thermal partition sum, is given by
\begin{equation} \label{eq:sm6}
\delta^L_\tau = \sum_{x=0}^{L-1}\left[f^L_\tau(x)-\frac{1}{L}\right],
\end{equation}
and the part of $\delta^L_\tau$ that is bound in the localized term of \eq{eq:sm5} can be estimated by
\begin{equation} \label{eq:sm7}
\delta^L_{\tau,loc.} \simeq \delta^L_\tau - Lf^L_\tau(L/2)
\end{equation}
via substracting the delocalized tail. The difference in the portion of additional localized weight for quench and equilibrium can thus be inspected by the imbalance
\begin{equation} \label{eq:sm8}
I(L) = \frac{\delta^L_{\mathrm{q.},loc.}}{\delta^L_{\mathrm{q.}}} - \frac{\delta^L_{\mathrm{eq.},loc.}}{\delta^L_{\mathrm{eq.}}},
\end{equation}
which is shown in \fig{fig:7} (inset) and goes to a finite value for $L\rightarrow\infty$, which expresses the increased localization in the long time integrated quench.

We emphasize that a crucial aspect leading to non-thermal behaviour is the single-particle nature of the effective model, which stems from the sole dynamical center present in the maximally staggered inital state, and is thus a direct consequence of the constrained nature of the dimer model.

\subsection{Thermodynamic Limit: Robustness of pyramid structures}
As mentioned in the main text, at vanishing energy density with $\xi/L\rightarrow const.$ as $L\rightarrow \infty$, the individual pyramids of the inital state can be treated independently. In order to show this we provide an argument for the existence of states of similar energies, which will thus contribute similarly to the thermodynamic partition sum, but that are seperated by an extensive amount of off-resonant plaquette-flip processes.

\begin{figure}[b]
\centering
\includegraphics[width=.95\columnwidth]{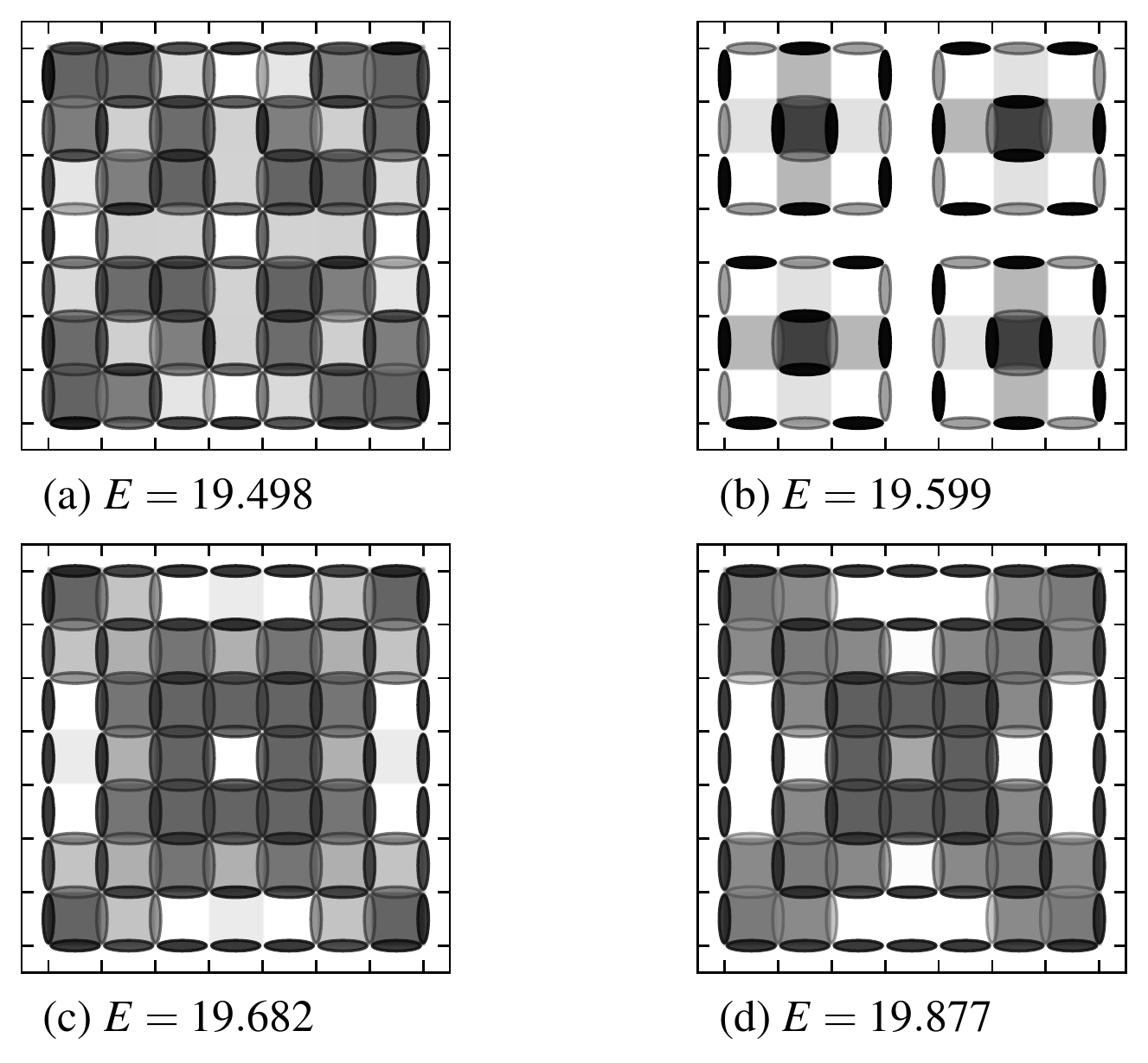}
\caption{\textbf{Localization at vanishing energy densities.} Illustration of both plaquette-density and dimer-density for rotationally invariant eigenstates on $L=8$ at large $V/J=5$. All states (a)-(d) are taken at energies $E$ dominated by the presence of about four parallel plaquettes. (a),(c),(d) clearly show the structure of a large pyramid with $\xi=8$ and the corresponding delocalization along the system diagonal as discussed previously. On the other hand, (b) is localized around pyramids of size $\xi=4$.}
\label{fig:8}
\end{figure}

For example, on a system of size $L\times L$, a state consisting of pyramids of fixed average length $\xi_1$, thus hosting $(L/\xi_1)^2$ active defects, is close in energy to a state with different length scale $\xi_2>\xi_1$, which in turn has already melted up $\xi_2^2/\xi_1^2-1$ plaquettes per pyramid (on average) in its interior. Starting from the former state, to reach the latter, we can again estimate the lowest order in perturbation theory that allows for local rearrangements. Via the same line of reasoning as in the main text for finite staggered scales, we obtain approximately
\begin{equation} \label{eq:sm9}
\left(\xi'\right)^2\sim\min\left\{L^2,\left[\xi_1^2/\left(1-\xi_1^2/\xi_2^2\right)\right]^2\right\}
\end{equation}
as the lowest order of perturbative rearrangements.
Since $\xi_1$ is of the order of the system size, at least $\mathcal{O}(L^2)$ consecutive plaquette flips that are off-resonant from the initial energy shell are required. This implies a matrix element between the two states that is bounded from above by $\Delta \lesssim J(J/V)^{c\,L^2}$, with $c\sim \min\left\{1,\xi_1^4/L^2\right\}$ of order $\mathcal{O}(1)$. On the other hand, the many-body level spacing that is relevant for the hybridization of two states is, in the absence of symmetries beyond $C_4$, bounded by the total number of dimer configurations, $\epsilon\gtrsim J\exp(-G/\pi L^2)$, where $G$ is the Catalan's constant \cite{Fisher_1961_Dimer}. The two states will then not be able to hybridize if their level spacing is larger than the matrix element between them, $\Delta \lesssim \epsilon$; hence, in this approximation for $\log(V/J)\gtrsim G/\pi c$. This argument directly shows that at $T=0$, the pyramid structure of the inital state will be preserved in the post-quench dynamics, implying the existence of non-thermal eigenstates just above the ground state.

\begin{figure}[t]
\centering
\includegraphics[width=.50\columnwidth]{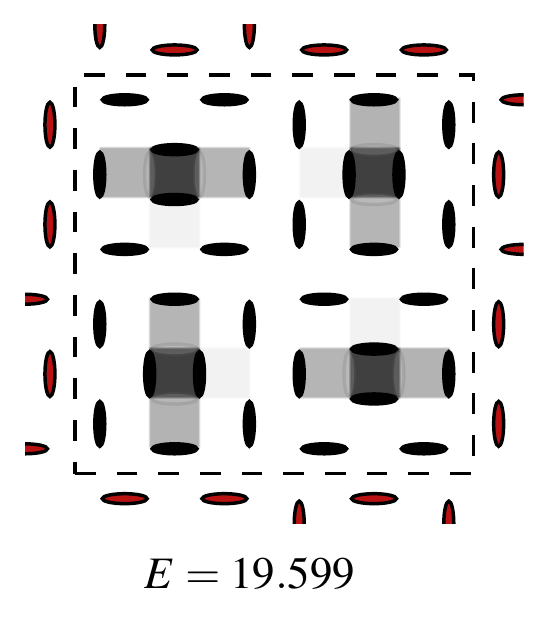}
\caption{\textbf{Staggered boundary conditions.} Dimers on the outer side of the boundary (marked in red) are fixed in a staggered configuration of scale $\xi=4$. The plaquette- and dimer-densities of the lowest eigenstate at $V/J=5$ demonstrate the continuation of $\xi=4$ on the inner side of the boundary.}
\label{fig:9}
\end{figure}

We further observe that according to \eq{eq:sm9}, thermal relaxation can only occur if $\xi^4$ is of order smaller than $\mathcal{O}(L^2)$. Therefore, non-thermal behaviour will persist in the thermodynamic limit even for an infinite amount of plaquettes in the inital state so long as
\begin{equation} \label{eq:sm10}
\lim_{L\rightarrow \infty}\;\xi\, L^{-1/2}\neq 0
\end{equation}
holds true, which still corresponds to vanishing energy densities.

To illustrate the above argument numerically at least within the bounds of finite size limitations, we can investigate the supposed rearrangement $\xi_1=4 \rightarrow \xi_2=8$ on $L=8$. \fig{fig:8} shows the potential energy landscape, as well as the dimer-density $\braket{\hat{n}_l}$ on the bonds $l$ of the lattice for selected eigenstates in the rotationally invariant sector at $V/J=5$. The eigenstates of \fig{fig:8} are adjacent in the spectrum of $\hat{H}$, chosen at an energy that corresponds approximately to the presence of $(L/\xi_1)^2=4$ plaquettes in the system. Consistent with the argument presented above, we find that states with different $\xi_1,\xi_2$ are \emph{not} hybridized.

We can extend these numerical considerations by changing from OBCs to fixed boundaries that correspond to a staggered background of $\xi=4$. In this scenario, we envisage the $8\times 8$ system as a patch of a larger system prepared at $\xi=4$. As shown in \fig{fig:9}, the \emph{lowest} energy eigenstate, with $\xi=4$, has not hybridized with any states of different $\xi$, consistent with our analytical formula \eq{eq:7} from the main text.

\end{document}